\documentclass[12pt]{article} \usepackage{amssymb,latexsym}
\usepackage{amsmath}
 
\columnwidth\textwidth

\begin{document}

\newtheorem{df}{Definition} \newtheorem{thm}{Theorem} \newtheorem{lem}{Lemma}
\newtheorem{rl}{Rule}
\begin{titlepage}
 
\noindent
 
\begin{center} {\LARGE Changes of separation status during registration and
scattering} \vspace{1cm}

P. H\'{a}j\'{\i}\v{c}ek\\ Institute for Theoretical Physics \\ University of
Bern \\ Sidlerstrasse 5, CH-3012 Bern, Switzerland \\ hajicek@itp.unibe.ch

\vspace{1cm}

December 2011 \\ \vspace{1cm}
 
 
\vspace*{2cm}
 
\nopagebreak[4]
 
\begin{abstract} In our previous work, a new approach to the notorious problem
of quantum measurement was proposed. Existing treatments of the problem were
incorrect because they ignored the disturbance of measurement by identical
particles and standard quantum mechanics had to be modified to obey the
cluster separability principle. The key tool was the notion of separation
status. Changes of separation status occur during preparations, registrations
and scattering on macroscopic targets. Standard quantum mechanics does not
provide any correct rules that would govern these changes. This gives us the
possibility to add new rules to quantum mechanics that would satisfy the
objectification requirement. The method of the present paper is to start from
the standard unitary evolution and then introduce minimal corrections. Several
representative examples of registration and particle scattering on macroscopic
targets are analysed case by case in order to see their common features. The
resulting general Rule of Separation Status Changes is stated in the
Conclusion.
\end{abstract}

\end{center}

\end{titlepage}

\section{Introduction} In several recent papers
\cite{HT,hajicek1,hajicek2,survey,hajicek3}, a new understanding of quantum
mechanics is proposed, called The Reformed Quantum Mechanics. While the
statistical character and non-locality remain unchanged and are considered as
facts of life, the main thrust of the reform is aimed at the emergence of
classical theories. This is hindered by the three notorious problems: those of
realist interpretation, of classical properties and of quantum
measurement. Objective properties of quantum systems are assumed there to be
those that are uniquely defined by preparation rather than values of
observables. Then, there are enough objective properties to view quantum
systems as physical objects and quantum mechanics becomes as objective as any
other physical theory \cite{HT,survey}. Classical properties are understood as
specific objective properties of high-entropy states of macroscopic quantum
systems (including Newtonian mechanics) and classical limit is a suitably
taken high-entropy limit \cite{hajicek1,survey}. The present paper is a
continuation of our work on the measurement \cite{hajicek2,survey,hajicek3}.

It is well known that the quantum theory of measurement is in an
unsatisfactory state \cite{BLM,Ghirardi}. In \cite{hajicek2} a new idea is
described: First, any quantum theory of measurement that disregards the
disturbance of registration due to identical particles is proved to be
wrong. Second, notions of a D-local observable and a separation status have
been introduced and shown how they help to eliminate this disturbance. The
reformed theory allows only those operators that satisfy a $D$-locality
condition to be observables. Preparations must separate the quantum systems
from the sea of identical particles giving them a non-trivial separation
status characterised by some non-empty domain $D$ of space. Third,
registration of microsystems must use detectors, in which then the separation
status becomes trivial ($D = \emptyset$) again. Let us emphasise that the
change of separation status is an objective property of the composite
system+apparatus, as it follows from our definition of objective properties in
\cite{HT,survey}.

The standard quantum mechanics turns out to be just a theory of isolated
systems ($D = {\mathbb R}^3$) and subsystems of other isolated systems ($D =
\emptyset$) allowing only two separation statuses. Hence, it must be supplied
by a theory of general separation statuses and an additional rule governing
changes of separation status. Finally, fourth, the freedom in the choice of
the additional rules allows us to satisfy the existing observational evidence,
in particular the objectification requirement. The rule describes an objective
process inside a macroscopic detector including a change of kinematic
description, a unitary evolution and a state reduction. For details, see
\cite{hajicek2,survey,hajicek3}.

The idea that standard quantum theory must be corrected because all standard
observables are global while no registration apparatus can control the whole
space has also been put forward by K. K. Wan and his collaborators \cite{wan}
and \cite{wanb}, Sec.\ 2.12. Let us explain briefly what is the difference in
the aims for which local observables are used in \cite{wan} and
\cite{hajicek2}. Wan's first main idea is to consider measurement as a
scattering process. The crucial property of most scattering states is that
they become spatially separated from each other in the asymptotic
region. Local observables have no correlation between such states. Then, some
asymptotic states, even if pure, are equivalent, with respect to the set of
all local observables, to mixed states. Wan's second main idea is to apply the
corresponding superselection approach to classical observables and measurement
problem. For a description of the superselection approach see \cite{BLM}.

Our first main idea is that local observables help to eliminate the
disturbance of measurement due to identical particles. To keep the theory
sufficiently general, we consider any measurement as a process inside a
bounded region of space (e.g., a {\em finite} laboratory) and so avoid
situating registrations in asymptotic regions. This leads to additional
requirements on preparation, which must provide a non-trivial separation
status for any registrable system and the only observables are then those
local ones that are associated with this separation status. The second main
idea is that new dynamical rules governing changes of separation status must
be added to standard quantum mechanics, and this provides a natural framework
for a direct reduction of state.

An example of an additional rule governing a change of separation status has
been described in \cite{hajicek2}. In the present paper, we generalise the
ideas of \cite{hajicek2} and give a systematic theory of such rules. Our
leading principles are 1.\ to preserve as much of standard quantum mechanics
as possible, 2.\ to satisfy the objectification requirement and 3.\ to have a
rule applicable to each process that contains changes of separation status in
an unambiguous and observer-independent way.

The plan of the paper is as follows. The first two sections bring some
material from the previous papers to make the present paper self-contained. In
particular, Sec.\ 2 recapitulates the quantum model of measurement due to
Beltrametti, Cassinelli and Lahti \cite{belt} and the first part of Sec. 3
summarises the existing theory of separation status. Sec.\ 3.2 introduces
important technical tools for study of separation status changes, that of {\em
separated systems} and that of {\em formal evolution}.

Using these tools, Sec.\ 4 extends the study of registration processes in
\cite{hajicek2}, where a single registration of a one-particle system in a
vector state by an arrangement of ideal detectors with fixed signals were
dealt with. Registrations of systems in non-vector states, by detectors with
flexible signals and release of particles from detectors are also
considered. Sec.\ 4.1.4 brings generalisation of the theory to registrations
of many-particle systems. It agrees with the correlations between different
detectors due to the Hanbury Brown and Twiss effect \cite{HBT} as well as with
those in Einstein, Podolski and Rosen experiment. Sec.\ 4.2 discusses the case
of non-ideal detectors and introduces the notion of approximate probability
reproducibility. Particle tracks in cloud chambers are explained in Sec.\ 4.3
as multiple registrations. Formulas of Sec.\ 4 represent different cases of
non-unitary evolution. Sec.\ 5 studies changes of separation status that occur
during scattering on macroscopic systems. A formula that governs these cases
is unitary and agrees with the predictions of standard quantum mechanics.

Finally, the account of different cases of separation status changes in Secs.\
4 and 5 allows us to exclude some ideas and an analysis of what all the cases
have in common suggests how a general rule could look like. The resulting Rule
of Separation Status Change is stated and discussed in the Conclusion.

\section{Beltrametti-Cassinelli-Lahti model} In this section, we are going to
recapitulate the well-known ideas on measurement that will be needed or
criticised later. We describe a quite general measurement process, which we
call Beltrametti-Cassinelli-Lahti (BCL) model \cite{belt}.

Let a discrete observable ${\mathsf O}$ of system ${\mathcal S}$ with Hilbert
space ${\mathbf H}$ be measured. Let $o_n$ be the eigenvalues and
$\{\phi_{nj}\} \subset {\mathbf H}$ be a complete orthonormal set of
eigenvectors of ${\mathsf O}$,
$$
{\mathsf O}\phi_{nj} = o_n \phi_{nj}\ .
$$
We assume that $ n = 1,\cdots,N$ so that there is only a finite number of
different eigenvalues $o_n$. This is justified by the fact that no real
registration apparatus can distinguish all elements of an infinite set from
each other. It can therefore measure only a function of a general observable
that maps the spectrum onto a finite set of real numbers. Our observable
${\mathsf O}$ is such a function. The restriction to discrete observables is
also valid for real measurements. The continuous eigenvalues can be grouped
into small intervals and orthonormal bases can be chosen in the corresponding
subspaces \cite{BLM}.

Let the registration apparatus\footnote{In our language, a measurement
consists of preparation and registration so that what Ref.\ \cite{BLM} often
calls "measurement" is our "registration".} be a quantum
system ${\mathcal A}$ with Hilbert space ${\mathbf H}_{\mathcal A}$ and an
observable ${\mathsf A}$. Let ${\mathsf A}$ be a non-degenerate, discrete
observable with the same eigenvalues $o_n$ and with orthonormal set of
eigenvectors $\psi_n$,
$$
{\mathsf A}\psi_n = o_n \psi_n\ ,
$$
with possible further eigenvectors and eigenvalues. ${\mathsf A}$ is the
so-called {\em pointer observable} \cite{BLM}.

Let the measurement start with the preparation of ${\mathcal S}$ in state
${\mathsf T}$ and the independent preparation of ${\mathcal A}$ in state
${\mathsf T}_{\mathcal A}$. The initial state of the composite system
${\mathcal S} + {\mathcal A}$ is thus ${\mathsf T}\otimes {\mathsf
T}_{\mathcal A}$.

Let ${\mathcal S}$ and ${\mathcal A}$ then interact for a finite time by the
so-called {\em measurement coupling} and let the resulting state be ${\mathsf
U}({\mathsf T}\otimes {\mathsf T}_{\mathcal A}){\mathsf U}^\dagger$, where
${\mathsf U}$ is a unitary transformation on ${\mathbf H} \otimes {\mathbf
H}_{\mathcal A}$. If ${\mathcal A}$ is to measure ${\mathsf O}$, the
probability of ${\mathcal A}$ being in state $\psi_n$ after the interaction
must be the same as the probability of eigenvalue $o_n$ being registered on
${\mathcal S}$ as given by Born rule. This is called {\em probability
reproducibility} \cite{BLM}.

The evolution of composite ${\mathcal S} + {\mathcal A}$ due to the
measurement coupling can be generalised to be non-unitary by introducing some
environment and allowing the system to be only approximately isolated
\cite{BLM}, but this would not change the subsequent results in an important
way.

Now, there is a theorem \cite{belt}:
\begin{thm} Let a measurement fulfil all assumptions and conditions listed
above. Then, for any initial vector state $\psi \in {\mathbf H}_{\mathcal A}$,
there is a set $\{\varphi_{nl}\} \subset {\mathbf H}$ satisfying the
orthogonality conditions
\begin{equation}\label{orth} \langle \varphi_{nl}|\varphi_{nj}\rangle =
\delta_{lj}
\end{equation} such that ${\mathsf U}$ is a unitary extension of the map
\begin{equation}\label{unitar} \phi_{nl}\otimes \psi \mapsto
\varphi_{nl}\otimes \psi_n\ .
\end{equation}
\end{thm}

An observational evidence of many years of quantum experimenting is that each
individual measurement process leads to a definite result shown by the
apparatus. More precisely, the apparatus must be in one of the states
$|\psi_n\rangle\langle\psi_n|$ after each individual registration. This is
called {\em objectification requirement} \cite{BLM}.

Suppose that the initial state of ${\mathcal S}$ is an arbitrary vector state,
${\mathsf T} =|\phi\rangle\langle\phi|$. Decomposing $\phi$ into the
eigenstates,
\begin{equation}\label{decom} \phi = \sum_{nl} c_{nl}\phi_{nl}\ ,
\end{equation} we obtain from Eq.\ (\ref{unitar}) and the linearity of
${\mathsf U}$:
\begin{equation}\label{finalSA} \Phi_{\text{end}} = {\mathsf U} (\phi \otimes
\psi) = \sum_n \sqrt{p_n}\varphi^1_n\otimes \psi_n\ ,
\end{equation} where
\begin{equation}\label{Phik} \varphi^1_n = \frac{\sum_k
c_{nk}\varphi_{nk}}{\sqrt{\langle \sum_l c_{nl}\varphi_{nl}|\sum_j
c_{nj}\varphi_{nj}\rangle}}
\end{equation} and
$$
p_n = \sum_l |c_{nl}|^2
$$
is the probability that a registration of ${\mathsf O}$ performed on vector
state $\phi$ gives value $o_k$.

Eq.\ (\ref{unitar}) implies that the probability of value $o_n$ is $p_n$ if
${\mathsf 1} \otimes {\mathsf A}$ is registered on ${\mathcal S} + {\mathcal
A}$ in final state (\ref{finalSA}). Hence, apparatus $\mathcal A$ measures
observable ${\mathsf O}$. However, this final state is a linear superposition
of composite states containing different states $\psi_n$ of the
apparatus. This means that the apparatus is simultaneously in all states
$\psi_n$ for which coefficients $p_n$ are non-zero. For more discussion, see
\cite{BLM,hajicek2}.

Hence, the objectification requirement is not satisfied and the prediction of
the theoretical model does not agree with observation. This is called
"problem of quantum measurement" or "problem of
objectification", John Bell has called it the problem of "and"
versus "or" and Schr\"{o}dinger invented
"Schr\"{o}dinger cat" to visualise it in a provocative way. Von
Neumann introduced changes into quantum mechanics, the so-called "first
kind of dynamics", which was later called "the collapse of wave
function", to obtain agreement with observation. A modified version of
von Neumann's solution is adopted in our theory \cite{hajicek2}.

\section{Separation status} In \cite{hajicek2}, quantum mechanics is reformed
so that the disturbance of measurements due to remote identical particles is
avoided and the cluster separability principle is satisfied. Let us briefly
recapitulate and further develop this reform. We shall work with
$Q$-representation in this section.

\subsection{Basic definitions and rules} First, a locality requirement on
operators is introduced:
\begin{df} Let $D \subset {\mathbb R}^{3}$ be open. Operator with kernel
$a(\vec{x};\vec{x}')$ is \underline{$D$-local} if
$$
\int d^3x'\, a(\vec{x};\vec{x}') f(\vec{x}') = \int d^3x\, a(\vec{x};\vec{x}')
f(\vec{x}) = 0\ ,
$$
for any test function $f$ vanishing in $D$.
\end{df} An equivalent definition has been given in \cite{wan} and
\cite{wanb}, Sec.\ 2.12. For the generalisation to composite system see
\cite{survey}. All self-adjoint $D$-local operators of a system ${\mathcal S}$
form an algebra that will be denoted by ${\mathbf A}[{\mathcal S}]_D$.

The key notion of our theory is:
\begin{df} Let ${\mathcal S}$ be a particle\footnote{Particles have wave
functions with three arguments, composite systems containing $N$ particles
those with $3N$ arguments.} and $D \subset {\mathbb R}^3$ an open set such
that:
\begin{itemize}
\item Registrations of any ${\mathsf A} \in {\mathbf A}[{\mathcal S}]_D$ lead
to average $\langle \psi(\vec{x})|{\mathsf A}\psi(\vec{x})\rangle$ for all
vector states $\psi(\vec{x})$ of ${\mathcal S}$.\footnote{It seems that this
can be generalised to approximate, $\pm \epsilon$ say, equality of the average
to the expression $\langle \psi(\vec{x})|{\mathsf A}\psi(\vec{x})\rangle$,
leading to generalised separation status denoted by $(D,\epsilon)$. The
corresponding reformulation of the theory will be published in another paper.}
\item ${\mathcal S}$ is prepared in state $\psi(\vec{x})$ that does not vanish
in $D$.
\end{itemize} Then ${\mathcal S}$ is said to have \underline{separation
status} $D$.
\end{df} Generalisation to composite ${\mathcal S}$ and its non-vector states
are given in \cite{survey}. As an example, consider $D$ in which all wave
functions of particles identical to ${\mathcal S}$ vanish. A separation status
is called {\em trivial} if $D = \emptyset$.

We require next that any preparation of ${\mathcal S}$ must give it a
non-trivial separation status. Then elements of ${\mathbf A}[{\mathcal S}]_D$
are observables of ${\mathcal S}$ and \underline{only these are}. Standard
quantum mechanics assumes that all self-adjoint operators on the Hilbert space
of ${\mathcal S}$ are observables and can in principle be registered. This is
different in the reformed quantum mechanics: only some subset of all
self-adjoint operators contains observables and the subset even depends on
preparation. Next, exceptions to the standard rule on composition of identical
systems are described \cite{hajicek3}:
\begin{rl} Let $\mathcal S$ be prepared in state ${\mathsf T}$ and have
separation status $D \neq \emptyset$. Then its observables are elements of
${\mathbf A}[{\mathcal S}]_D$ and its state is ${\mathsf T}$ independently of
any remote system identical to $\mathcal S$ that may exist.
\end{rl} Composition of such states and observables satisfy
\begin{rl} ${\mathcal S}_1$ and ${\mathcal S}_2$ prepared in states ${\mathsf
T}_1$ and ${\mathsf T}_2$ with non-trivial separation statuses $D_1$ and
$D_2$, $D_1 \cap D_2 = \emptyset$. Then ${\mathcal S}_1 + {\mathcal S}_2$ has
state ${\mathsf T}_1 \otimes {\mathsf T}_2$ and its observables are elements
of ${\mathbf A}[{\mathcal S}_1]_{D_1} \otimes {\mathbf A}[{\mathcal
S}_2]_{D_2}$. This holds even if ${\mathcal S}_1$ and ${\mathcal S}_2$ have
particles of the same kind in common.
\end{rl}

For registrations, it is assumed:
\begin{rl} Any registration apparatus for \underline{microsystems} must
contain at least one detector and every "reading of a pointer value"
is a signal from a detector.
\end{rl} What constitutes a detector and its signal may be defined by detector
classifications, such as \cite{leo,stefan}. Rule 3 has many surprising
consequences: e.g., a generalisation of separation status \cite{hajicek2}. The
most important consequence is that by entering the sensitive matter of a
detector, a system is transferred from a non-trivial separation status into
the trivial one. For more discussion, see \cite{hajicek2,survey}.

It follows that preparation and registration acquire an additional importance
in the reformed quantum mechanics: they necessarily include changes of
separation status.

To summarize: Standard quantum mechanics is just a theory of isolated systems
and as such it is incomplete:
\begin{enumerate}
\item It admits only two separation statuses for any system ${\mathcal S}$:
\begin{enumerate}
\item ${\mathcal S}$ is isolated. Then $D = {\mathbb R}^3$ and all s.a.\
operators are observables.
\item ${\mathcal S}$ is a member of an isolated system containing particles
identical to ${\mathcal S}$. Then $D = \emptyset$ and there are no observables
for ${\mathcal S}$.
\end{enumerate}
\item It ignores the existence of separation-status changes and the fact that
such processes are objectively different from all other ones. Rules for
changes of separation status that can be derived from standard quantum
mechanics do not agree with observations in most cases.
\end{enumerate}

This seems to give us an opportunity to introduce new rules that govern
processes in which separation status changes. The conditions on such new rules
are:
\begin{enumerate}
\item They do not contradict the rest of quantum mechanics. That is, all
correct results of standard quantum mechanics remain valid.
\item They agree with, and explain, observational facts, in particularly the
objectification requirement.
\item They can be applied to any change of separation status in an unambiguous
and observer independent way.
\end{enumerate} A possible new rule for Beltrametti-Cassinelli-Lahti model
modified by ionisation gas detectors was proposed in \cite{hajicek2}. Points
1.\ and 2.\ were then satisfied but point 3. were not.

\subsection{Some technical tools} This subsection introduces some technical
tools to deal with separation status chang\-es. Notions of separated systems
and formal evolution will be introduced and their importance for analysis of
separation status changes explained.

We start with separated systems. Let ${\mathcal S}$ and ${\mathcal S}'$ be
systems with Hilbert spaces ${\mathbf H}$ and ${\mathbf H}'$. The composite
${\mathcal S} + {\mathcal S}'$ can be uniquely decomposed into subsystems,
$$
{\mathcal S} + {\mathcal S}' = {\mathcal B}_1 + \cdots + {\mathcal B}_b +
{\mathcal F}_1 + \cdots + {\mathcal F}_f\ ,
$$
so that ${\mathcal B}_n$ contains only bosons of the same kind for each $n =
1,\cdots,b$ and ${\mathcal F}_n$ contains only fermions of the same kind for
each $n = 1,\cdots,f$. The map
$$
{\mathsf P}_{as} : {\mathbf H} \otimes {\mathbf H}' \mapsto {\mathbf H}_{as}
$$
is defined as the orthogonal projection onto ${\mathbf H}_{as}$, where
${\mathbf H}_{as}$ is the representation subspace of a permutation group
representation on ${\mathbf H} \otimes {\mathbf H}'$, the elements of which
are symmetric over each set ${\mathcal B}_n$ and antisymmetric over each set
${\mathcal F}_n$. Thus, ${\mathsf P}_{as}$ is linear and self-adjoint.
\par \vspace*{.4cm} \noindent {\bf Example} Let ${\mathcal S}$ be a fermion
particle and ${\mathcal S}'$ a composite of one fermion of the same kind as
${\mathcal S}$ and some particle of a different kind. Let $\phi(\vec{x}_1)$ be
an element of ${\mathbf H}$ and $\phi'(\vec{x}_2,\vec{x}_3)$ that of ${\mathbf
H}'$, $\vec{x}_2$ being the coordinate of the fermion. Then
$$
\Psi(\vec{x}_1,\vec{x}_2,\vec{x}_3) = {\mathsf
P}_{as}\big(\phi(\vec{x}_1)\phi'(\vec{x}_2,\vec{x}_3)\big) =
\frac{1}{2}\big(\phi(\vec{x}_1)\phi'(\vec{x}_2,\vec{x}_3) -
\phi(\vec{x}_2)\phi'(\vec{x}_1,\vec{x}_3)\big)\ .
$$
\par \vspace*{.4cm} \noindent In general, ${\mathsf P}_{as}$ is non-invertible
and does not preserve norm. Another map we shall need is the normalisation,
$$
{\mathsf N} : {\mathbf H}_{as} \setminus \{0\} \mapsto {\mathbf H}_{as}\ ,
$$
which is, in general, neither linear nor invertible. Its range is the unit
sphere in ${\mathbf H}_{as}$.

We show that the maps are invertible in a special case of separation statuses.
\begin{df} Let composite ${\mathcal S} + {\mathcal S}'$ be prepared in state
$\bar{\mathsf T}$. Let ${\mathsf T} = tr_{{\mathcal S}'}[\bar{\mathsf T}]$ and
${\mathsf T}' = tr_{{\mathcal S}}[\bar{\mathsf T}]$ be states of ${\mathcal
S}$ and ${\mathcal S}'$ with separation statuses $D$ and $D'$, respectively,
satisfying
\begin{equation}\label{disjoint} D \cap D' = \emptyset\ .
\end{equation} Then ${\mathcal S}$ and ${\mathcal S}'$ are called separated.
\end{df} We limit ourselves to the non-entangled case, $\bar{\mathsf T} =
{\mathsf T} \otimes {\mathsf T}'$.

Consider first vector states $\phi$ and $\phi'$. Let us define map ${\mathsf
J}$ by
\begin{equation}\label{operj} {\mathsf J} = {\mathsf N}|_{{\mathbf H}_{as}
\setminus \{0\}} \circ {\mathsf P}_{as}\ ,
\end{equation} and let
$$
\Phi_{as} = {\mathsf P}_{as} (\phi \otimes \phi')\ ,\quad \Phi_{asn} =
{\mathsf J}(\phi \otimes \phi')\ .
$$
If ${\mathcal S}$ and ${\mathcal S}'$ are separated, then $\phi$ and $\phi'$
satisfy:
$$
\int d^3 x_i f'(\vec{x}_i)\phi(\vec{x}_1,\ldots,\vec{x}_K) = 0
$$
for any $i = 1,\ldots,K$ and for any test function $f'$ with $\text{supp} f'
\subset D'$, and
$$
\int d^3 x_i f(\vec{x}_i)\phi'(\vec{x}_1,\ldots,\vec{x}_L) = 0
$$
for any $i = 1,\ldots,L$ and for any test function $f$ with $\text{supp} f
\subset D$.

Let $f'$ be a test function such that $f' \in {\mathbf H}'$ with $\text{supp}
f \subset (D'\times)^L$, where $(D'\times)^L$ is an abbreviation for the
Cartesian product of $L$ factors $D'$. Let us define map $R[f',D'] : {\mathbf
H}_{as} \mapsto {\mathbf H}$ by
\begin{multline*} (R[f',D'] \Phi_{as})(\vec{x}_1,\ldots,\vec{x}_K) \\ = \int
d^3 x_{K+1}\ldots d^3 x_{K+L}
f'(\vec{x}_{K+1},\ldots,\vec{x}_{K+L})\Phi_{as}(\vec{x}_1,\ldots,\vec{x}_K,\vec{x}_{K+1},\ldots,\vec{x}_{K+L})\
,
\end{multline*} and similarly, for test function $f \in {\mathbf H}$ and
$\text{supp} f \subset (D\times)^K$, $R[f,D] : {\mathbf H}_{as} \mapsto
{\mathbf H}'$ by
\begin{multline*} (R[f,D] \Phi_{as})(\vec{x}_{K+1},\ldots,\vec{x}_{K+L}) \\ =
\int d^3 x_1\ldots d^3 x_K
f(\vec{x}_1,\ldots,\vec{x}_K)\Phi_{as}(\vec{x}_1,\ldots,\vec{x}_K,\vec{x}_{K+1},\ldots,\vec{x}_{K+L})\
.
\end{multline*} Then, we obtain easily:
$$
R[f',D']\Phi_{as} = N_f' \phi(\vec{x}_1,\ldots,\vec{x}_K)\ ,
$$
where
$$
N_f' = N_{as} \int d^3 x_{K+1}\ldots d^3
x_{K+L}f'(\vec{x}_{K+1},\ldots,\vec{x}_{K+L})\phi'(\vec{x}_{K+1},\ldots,\vec{x}_{K+L})\
,
$$
and $N_{as}$ is the normalisation factor defined by $P_{as}$. $N_f'$ is
non-zero for at least some $f'$. Similarly,
$$
R[f,D]\Phi_{as} = N_f \phi'(\vec{x}_{K+1},\ldots,\vec{x}_{K+L})\ ,
$$
where
$$
N_f = N_{as} \int d^3 x_1\ldots d^3 x_K f(\vec{x}_1,\ldots,\vec{x}_K)
\phi(\vec{x}_1,\ldots,\vec{x}_K) \ .
$$
Thus, we obtain both functions $\phi(\vec{x}_1,\ldots,\vec{x}_K)$ and
$\phi'(\vec{x}_{K+1},\ldots,\vec{x}_{K+L})$ up to normalisation. As the
functions are normalised, they can be reconstructed. Analogous steps work for
$\Phi_{asn}$.

For the generalisation of these ideas to state operators, we shall need
adjoints of operators $R[f',D']$ and $R[f,D]$. The definition of
$R[f,D']^\dagger : {\mathbf H} \mapsto {\mathbf H}_{as}$ is
$$
(R[f',D']^\dagger \phi,\Phi) = (\phi,R[f',D']\Phi)
$$
and simple calculation yields
$$
R[f',D']^\dagger \phi = {\mathsf P}_{as}(\phi \otimes f^{\prime *})\ .
$$
Similarly,
$$
R[f,D]^\dagger \phi' = {\mathsf P}_{as}(f^* \otimes \phi')\ .
$$

Map ${\mathsf J}$ can be generalised to tensor product of any two operators
${\mathsf T}$ of $\mathcal S$ and ${\mathsf T}'$ of ${\mathcal S}'$. Map
${\mathsf T} \otimes {\mathsf T}' \mapsto {\mathsf P}_{as}({\mathsf T} \otimes
{\mathsf T}'){\mathsf P}_{as}$ is linear in both ${\mathsf T}$ and ${\mathsf
T}'$ and its result is an operator on ${\mathbf H} \otimes {\mathbf H}'$ that
leaves ${\mathbf H}_{as}$ invariant. Operator ${\mathsf P}_{as}({\mathsf T}
\otimes {\mathsf T}'){\mathsf P}_{as} : {\mathbf H}_{as} \mapsto {\mathbf
H}_{as}$ is self-adjoint and positive if ${\mathsf T}$ and ${\mathsf T}'$ are
state operators. Let $\{\psi_n\}$ be a basis ${\mathbf H}$ and
$\{\psi'_\alpha\}$ that of ${\mathbf H}'$. We can write
$$
{\mathsf T} = \sum_{mn} T_{mn}|\psi_m\rangle\langle\psi_n|\ ,\quad {\mathsf
T}' = \sum_{\alpha\beta} T'_{\alpha\beta}
|\psi'_\alpha\rangle\langle\psi'_\beta|\ .
$$
Then
$$
{\mathsf P}_{as}({\mathsf T} \otimes {\mathsf T}'){\mathsf P}_{as} =
\sum_{mn}\sum_{\alpha\beta}T_{mn}T'_{\alpha\beta} |{\mathsf
P}_{as}(\psi_m\otimes \psi'_\alpha)\rangle \langle{\mathsf P}_{as}(\psi_n
\otimes \psi'_\beta)|\ .
$$
Operator ${\mathsf P}_{as}({\mathsf T} \otimes {\mathsf T}'){\mathsf P}_{as}$
is not normalised even if ${\mathsf T}$ and ${\mathsf T}'$ are state
operators. Let us define
$$
{\mathsf J}({\mathsf T} \otimes {\mathsf T}') = \frac{{\mathsf
P}_{as}({\mathsf T} \otimes {\mathsf T}'){\mathsf P}_{as}}{tr[{\mathsf
P}_{as}({\mathsf T} \otimes {\mathsf T}'){\mathsf P}_{as}]} \ .
$$
Clearly, ${\mathsf J}$ maps states on states. Now, the above proof that vector
states $\phi$ and $\phi'$ can be reconstructed from ${\mathsf J}(\phi \otimes
\phi')$ can be easily extended to general states ${\mathsf T}$ and ${\mathsf
T}'$.

Moreover, for separated systems, the "individual" observables from
${\mathbf A}[{\mathcal S}]_D$ and ${\mathbf A}[{\mathcal S}']_{D'}$ can be
recovered from operators on ${\mathbf H}_{as}$ that are, in turn, constructed
from operators either of ${\mathbf A}[{\mathcal S}]_D$ or of ${\mathbf
A}[{\mathcal S}']_{D'}$.

For instance, consider systems and states defined in the above Example and let
${\mathsf a} \in {\mathbf A}[{\mathcal S}]_D$. Then ${\mathsf A}$ constructed
from ${\mathsf a}$ is an operator on ${\mathbf H}_{as}$ that is defined by its
kernel
$$
a(\vec{x}_1;\vec{x}'_1) \delta(\vec{x}_2 - \vec{x}'_2) \delta(\vec{x}_3 -
\vec{x}'_3) + a(\vec{x}_2;\vec{x}'_2) \delta(\vec{x}_1 - \vec{x}'_1)
\delta(\vec{x}_3 - \vec{x}'_3)
$$
so that
$$
({\mathsf A}\Psi)(\vec{x}_1,\vec{x}_2,\vec{x}_3) = \frac{1}{2}\big(({\mathsf
a}\phi)(\vec{x}_1)\phi'(\vec{x}_2,\vec{x}_3) - ({\mathsf
a}\phi)(\vec{x}_2)\phi'(\vec{x}_1,\vec{x}_3)\big)\ .
$$
Then,
$$
R[f',D']({\mathsf A}\Psi) = N_f' ({\mathsf a}\phi)(\vec{x}_1)\ ,
$$
where
$$
N_f' = \frac{1}{2} \int d^3 x_2d^3 x_3 f'(\vec{x}_2,\vec{x}_3)
\phi'(\vec{x}_2,\vec{x}_3)\ .
$$
But $\phi(\vec{x}_1)$, $\phi'(\vec{x}_2,\vec{x}_3)$ and
$f'(\vec{x}_2,\vec{x}_3)$ are known, hence, as $\phi$ is arbitrary, ${\mathsf
a}$ is well-defined.

To summarise: for separated systems ${\mathcal S}$ and ${\mathcal S}'$, there
are two equivalent descriptions: the {\em standard QM description} of
${\mathcal S} + {\mathcal S}'$ on the Hilbert space ${\mathbf H}_{as}$ and the
{\em reformed QM description} on ${\mathbf H} \otimes {\mathbf H}'$ explained
in Sec.\ 3.1.

Now, we come to the notion of formal evolution.
\begin{df} Let system ${\mathcal S}$ be initially ($t = t_1$) prepared in
state ${\mathsf T}$ and simultaneously another quantum system ${\mathcal S}'$
in state ${\mathsf T}'$. Let composite ${\mathcal S} + {\mathcal S}'$ be
isolated and have a time-independent Hamiltonian defining a unitary group
${\mathsf U}(t-t_1)$ of evolution operators on ${\mathbf H}_{as}$. Then, the
standard quantum mechanical evolution of ${\mathcal S} + {\mathcal S}'$,
\begin{equation}\label{sqme} {\mathsf T}_{\mathsf J}(t) = {\mathsf
U}(t-t_1){\mathsf J}\Big({\mathsf T} \otimes {\mathsf T}'\Big){\mathsf
U}(t-t_1)^\dagger\ ,
\end{equation} is called {\em formal evolution} of two interacting systems
${\mathcal S}$ and ${\mathcal S}'$.
\end{df} The choice of "initial state" ${\mathsf J}({\mathsf T}
\otimes {\mathsf T}')$ for the formal evolution clearly contradicts Rules 1
and 2. The change of kinematic description that accompanies change of
separation status \cite{hajicek2} has been arbitrarily shifted into the
past. This is why the evolution is called "formal". The name is also
justified by the fact that this evolution does not agree with observation in
many cases of separation status change. In our reformed quantum mechanics, we
have to define physical evolution in a different way. However, the formal
evolution is our first step in the mathematical analysis of separation status
changes. With its help, we can even recognise that a change of separation
status has taken place. For example:

Let ${\mathcal S}$ and ${\mathcal S}'$ be two quantum systems, ${\mathcal S}$
containing $K$ particles and ${\mathcal S}'$ containing $L$ particles. Let the
systems be prepared, at time $t_1$, in states ${\mathsf T}$ and ${\mathsf T}'$
with separation statuses $D_1$ and $D'$, respectively, so that $D_1 \cap D' =
\emptyset$. Let the formal evolution of the composite ${\mathcal S} +
{\mathcal S}'$ for the initial state ${\mathsf T}_{\mathsf J}(t_1) = {\mathsf
J}({\mathsf T} \otimes {\mathsf T}')$ be described by its kernel in
$Q$-representation:
$$
T_{\mathsf J}(t)(\vec{x}_1,\ldots,\vec{x}_K,
\vec{x}_{K+1},\ldots,\vec{x}_{K+L};
\vec{x}'_1,\ldots,\vec{x}'_K,\vec{x}'_{K+1},\ldots,\vec{x}'_{K+L})\ .
$$
\begin{enumerate}
\item Suppose that, for some $t_2 > t_1$, $\text{supp}\ T_{\mathsf J}(t_2) =
(D'\times)^{2(K+L)}$. Then we can say: at time $t_2$, the separation status of
${\mathcal S}$ is $\emptyset$, that of ${\mathcal S}'$ is $D'$ and that of the
composite ${\mathcal S} + {\mathcal S}'$ is also $D'$ or, that ${\mathcal S}$
is {\em swallowed} by ${\mathcal S}'$.\footnote{This can easily be generalised
to a more realistic condition, e.g., $\int_{(D'\times)^{K+L}}d^3x_1\ldots
d^3x_{K+L}T_{\mathsf
J}(t_2)(\vec{x}_1,\ldots,\vec{x}_{K+L};\vec{x}_1,\ldots,\vec{x}_{K+L}) \approx
1$.}
\item Suppose that, for some $t_3 > t_2$, there is a set $D_3$, $D_3 \cap D' =
\emptyset$, such that the kernel $T_{\mathsf J}(t_3)$ has the properties:
\begin{enumerate}
\item For any test function $f' \in {\mathbf H}'$ and
$$
\text{supp}f' = (D'\times)^L\ ,\quad R[f',D']T_{\mathsf
J}(t_3)R[f',D']^\dagger \neq 0\ ,
$$
${\mathsf N}(R[f',D']T_{\mathsf J}(t_3)R[f',D']^\dagger)$ is a state of
${\mathcal S}$ independent of $f'$.
\item For any test function $f \in {\mathbf H}$ and
$$
\text{supp}f = (D_3 \times)^K\ ,\quad R[f,D_3]T_{\mathsf
J}(t_3)R[f,D_3]^\dagger \neq 0\ ,
$$
${\mathsf N}(R[f,D_3]T_{\mathsf J}(t_3)R[f,D_3]^\dagger)$ is a state of
${\mathcal S}'$ independent of $f$.
\item For any test function $g \in {\mathbf H}$ and $\text{supp}g = (D_3
\times)^K$, we have
$$
{\mathsf N}(R[f',D']T_{\mathsf J}(t_3)R[f',D']^\dagger)|g\rangle = 0\ .
$$
\item For any test function $g' \in {\mathbf H}'$ and $\text{supp}g' = (D'
\times)^L$, we have
$$
{\mathsf N}(R[f,D_3]T_{\mathsf J}(t_3)R[f,D_3]^\dagger)|g'\rangle = 0\ .
$$
\end{enumerate} Then we can say: the systems become separated again at time
$t_3 > t_2$, system ${\mathcal S}$ being in state ${\mathsf
N}(R[f',D']T_{\mathsf J}(t_3)R[f',D']^\dagger)$ with separation status $D_3$
and system ${\mathcal S}'$ in state ${\mathsf N}(R[f,D_3]T_{\mathsf
J}(t_3)R[f,D_3]^\dagger)$ with separation status $D'$.
\end{enumerate}

Although we can find the separation statuses of ${\mathcal S}$ and ${\mathcal
S}'$ by studying the formal evolution of ${\mathcal S} +{\mathcal S}'$, we
cannot claim that this is the physical evolution of the composite. Hence, the
next question is how the formal evolution is to be corrected in the case that
it leads to separation-status changes. This will be studied in the next
section.

\section{Reformed models of registration} We shall now analyse several cases
of registration and try the minimal modifications of standard quantum
mechanics so that the objectification requirement could still be
satisfied. The modification starts by introducing detectors, the formal
evolution and some phenomenological model assumptions analogous to BCL method.

\subsection{Ideal detectors} First, we simplify things by assumption that the
detectors are ideal. For an ideal detector, the number of events registered by
the detector equals the number of events impinging on it ({\em intrinsic
efficiency} equal to 1). We also restrict ourselves to the sensitive matter of
the detector, denote it by ${\mathcal A}$ and speak of it as of the
detector. Initially, the registered system ${\mathcal S}$ and ${\mathcal A}$
are separated. We can, therefore, speak of initial states $\phi_{mk}$ of
${\mathcal S}$ as in Sec.\ 2 and ${\mathsf T}$ of ${\mathcal A}$, where
${\mathsf T}$ is assumed to be a stationary, high entropy state.

A {\em direct signal} of ${\mathcal A}$ is the macroscopic signal available
from the sensitive matter of the detector. Its possible transformation into an
electronic signal (such as for scintillation detectors) and further
amplification by an electronic amplifier connected to the detector is not
included in it. If we speak about detector signals, we always mean the direct
ones.

Eq.\ (\ref{unitar}) has now to be replaced by the formal evolution of
${\mathcal S} + {\mathcal A}$ on ${\mathsf H}_{as} = {\mathsf P}_{as}({\mathsf
H} \otimes {\mathsf H}_{\mathcal A})$. Let us write a suitable initial state
as follows:
\begin{equation}\label{init} {\mathsf T}_{\text{init}}({\mathrm c}) = {\mathsf
J}\left(\sum_{kl} c_k c^*_l |\phi_{mk}\rangle\langle \phi_{ml}| \otimes
{\mathsf T}\right) = {\mathsf N}\left(\sum_{kl} c_k c^*_l{\mathsf
P}_{as}(|\phi_{mk}\rangle\langle \phi_{ml}| \otimes {\mathsf T}){\mathsf
P}_{as}\right)\ ,
\end{equation} where $c_k$ are components of a unit complex vector ${\mathrm
c}$. Its evolution by any unitary map ${\mathsf U}$ is
\begin{equation}\label{NUP} {\mathsf U}{\mathsf T}_{\text{init}}({\mathrm
c}){\mathsf U}^\dagger = {\mathsf N}\left(\sum_{kl} c_k c^*_l{\mathsf
U}{\mathsf P}_{as}(|\phi_{mk}\rangle\langle \phi_{ml}| \otimes {\mathsf
T}){\mathsf P}_{as}{\mathsf U}^\dagger\right)\ .
\end{equation} It is, therefore, sufficient to consider operators ${\mathsf
P}_{as}(|\phi_{mk}\rangle \langle\phi_{ml}| \otimes {\mathsf T}){\mathsf
P}_{as}$ and their evolution for different possible values of $m$, $k$ and
$l$.

Let the formal evolution on ${\mathsf H}_{as}$ between the initial and an end
state be given by unitary map ${\mathsf U}$. It defines operators ${\mathsf
T}'_{mkl}$ on ${\mathsf H}_{as}$:
\begin{equation}\label{fulev} {\mathsf U}{\mathsf P}_{as}(|\phi_{mk}\rangle
\langle\phi_{ml}| \otimes {\mathsf T}){\mathsf P}_{as} {\mathsf U}^\dagger =
N{\mathsf T}'_{mkl}
\end{equation} where $N$ is a normalisation constant due to map ${\mathsf
P}_{as}$ not preserving norms. It is chosen so that
$$
tr[{\mathsf T}'_{mkl}] =\delta_{kl}\ .
$$
Let us formulate our model assumptions in terms of operators ${\mathsf
T}'_{mkl}$.
\begin{description}
\item[A] For any complex unit vector ${\mathrm c}$, state $\sum_{kl} c_k c^*_l
{\mathsf T}'_{mkl}$ includes a direct signal of the detector.
\item[B] For any pair of complex unit vectors ${\mathrm c}$ and ${\mathrm
c}'$, the states $\sum_{kl} c_k c^*_l {\mathsf T}'_{mkl}$ and $\sum_{kl} c'_k
c'^*_l {\mathsf T}'_{mkl}$ are not macroscopically different. That is, the
signal of the detector depends only on $m$ so that the detector registers
${\mathsf O}$.
\item[C] For any complex unit vector ${\mathrm c}$, state $\sum_{kl} c_k c^*_l
{\mathsf T}'_{mkl}$ describes system ${\mathcal S}$ being swallowed by
${\mathcal A}$, that is, the separation status of ${\mathcal S}$
changes. Hence, we cannot reproduce any particular state operator on ${\mathbf
H}$ as an end state of ${\mathcal S}$ and on ${\mathbf H}_{\mathcal A}$ as an
end state of ${\mathcal A}$. In general, it is not true that ${\mathcal S}$
and ${\mathcal A}$ are each in a well-defined state at the end.
\end{description}

If the formal evolution were applied to general initial state $\phi$ of
${\mathcal S}$ with decomposition (\ref{decom}) then the end state of the
composite ${\mathcal S} + {\mathcal A}$ would contain linear superposition of
different detector signals and the objectification requirement would be
violated. We shall therefore try next to weaken the assumption of unitarity.

\subsubsection{Flexible-signal detectors} Detectors can be divided in {\em
fixed-signal} and {\em flexible-signal} ones. For a fixed-signal detector, the
amplification erases differences of states $\sum_{kl} c_k c^*_l {\mathsf
T}'_{mkl}$ so that the signal is independent not only of ${\mathrm c}$ but
also of $m$. An example is a Geiger-Mueller counter. A flexible-signal
detector, such as a proportional counter, gives different signals for
different $m$.

The minimal change of the unitarity assumption results from the consequence of
assumptions A and B that the formal evolution of initial states $\phi$
constructed from all eigenstates of ${\mathcal S}$ with one fixed eigenvalue,
$$
\phi = \sum_k c_k \phi_{mk}\ ,
$$
does not lead to violation of objectification requirement.

Let us call this part of formal evolution a {\em channel} or {\em $m$-th
channel}. For a general initial state $\phi$, decomposition (\ref{decom}) can
be written as
$$
\phi = \sum_m \sqrt{p_m} \frac{c_{mk}}{\sqrt{p_m}}\ \phi_{mk}
$$
and $(\sqrt{p_m})^{-1}c_{mk}$ is a complex unit vector. Thus, $\phi$ is now a
linear superposition of different channels and we have to put all channels
together so that the result agrees with the objectification requirement. The
unique possibility is:
\begin{equation}\label{endflex} {\mathsf T}_{flex} = \left(\sum_{m=1}^N\right)
p_m \sum_{kl} \frac{c_{mk}c^*_{ml}}{p_m} {\mathsf T}'_{mkl}\ .
\end{equation} End state (\ref{endflex}) has the form of a convex linear
combination of states of the composite ${\mathcal S} + {\mathcal A}$, each of
which includes only one detector signal, and the combination is the gemenge
structure of the end state. In general, such an additional {\em reduction} of
the end state to a non-trivial gemenge cannot be the result of a unitary
evolution. The formal evolution defines the channels but remains valid only
within each channel. Observe that this is sufficient to recognise whether the
separation status has changed or not. Moreover, we can accept the validity of
Eq.\ (\ref{fulev}) and all properties A--C of operators ${\mathsf T}'_{mkl}$
as model assumptions without requiring full unitarity.

\subsubsection{Fixed-signal detectors} This is the case considered in
\cite{hajicek2}. Let state $\phi$ of particle ${\mathcal S}$ be prepared with
separation status $D$. Let ${\mathcal S}$ be manipulated by fields and screens
in $D$ so that beams corresponding to different eigenvalues of ${\mathsf O}$
become spatially separated.

Let the detector ${\mathcal A}$ be an array of $N$ fixed-signal sub-detectors
${\mathcal A}^{(m)}$ prepared in initial states ${\mathsf T}^{(m)}$ with
separation statuses $D^{(m)}$ where $D^{(n)} \cap D^{(m)} = \emptyset$ for all
$n \neq m$ and $ D^{(m)} \cap D = \emptyset$ for all $m$. We assume further
that the sub-detectors are placed at the boundary of $D$ in such a way that
the beam corresponding to eigenvalue $o_m$ will impinge on sub-detector
${\mathcal A}^{(m)}$ for each $m$. Each sub-detector ${\mathcal A}^{(m)}$
interacts with ${\mathcal S}$ as a whole and processes running in different
sub-detectors do not influence each other.

It has been shown that every observable can in principle be registered by this
kind of measurement (see \cite{wanb}, Sec.\ 3.6). The definition feature of it
is that different eigenvalues of the observable are associated with disjoint
regions of space and its registration can then be reduced to that of
position. However, even if the objectification problem could be solved for
such registrations, it still remains unsolved for other kinds of registration
(such as that described in the previous section), which no doubts exist and
exhibit the objectification effect.

In general, ${\mathcal S}$ hits all sub-detectors simultaneously because it is
present in all beams simultaneously. However, ${\mathcal S}$ in initial state
$\sum_{kl}c_kc^*_l |\phi_{mk}\rangle\langle\phi_{ml}|$ for any complex unit
vector ${\mathrm c}$ interacts only with sub-detector ${\mathcal
A}^{(m)}$. The formal evolution of ${\mathcal S} + {\mathcal A}^{(m)}$ can
then be decomposed into
$$
{\mathsf U}{\mathsf P}_{as}(|\phi_{mk}\rangle\langle\phi_{ml}| \otimes
{\mathsf T}^{(m)}){\mathsf P}_{as}{\mathsf U}^\dagger = N^{-1}{\mathsf
T}^{(m)\prime}_{kl}
$$
and we adopt assumptions 1--3 for operators ${\mathsf T}^{(m)\prime}_{kl}$.

Again, we have to put all channels together in the correct way. The end state
of ${\mathcal S} + {\mathcal A}$ for any initial state $\phi$ of ${\mathcal
S}$ then is
\begin{equation}\label{endfix} {\mathsf T}_{\text{fix}} =
\left(\sum_{m=1}^N\right) p_m \sum_{kl} \frac{c_{mk}c^*_{ml}}{p_m} {\mathsf
T}^{(m)\prime}_{kl} \otimes \prod_{r=1}^{N\setminus m}\otimes {\mathsf
T}^{(r)}\ ,
\end{equation} where $\prod_{r=1}^{N\setminus m}$ denotes the product of all
terms except for that with $r=m$ and coefficients $c_{mk}$ are defined by Eq.\
(\ref{decom}). Again, formula (\ref{endfix}) represents a non-trivial
reduction, where only the channels evolve unitarily.

\subsubsection{Some comments and generalisations} To discuss Eqs.\
(\ref{endflex}) and (\ref{endfix}), let us distinguish absorbing and
non-absorbing detectors \cite{hajicek2}. An absorbing detector does never
release a particle which it detects, a non-absorbing one always releases
it. We can consider only Eq.\ (\ref{endfix}), which will be needed later, the
other case is similar. If the detectors are absorbing, then state ${\mathsf
T}_{\text{fix}}$ evolves with ${\mathcal S}$ staying inside ${\mathcal
A}$. ${\mathcal S}$ is not manipulable and can be considered as lost in the
detector.

The case of non-absorbing detectors is more interesting. Extension of the
formal evolution in each channel then leads to separation of the two systems
at some later time. Further evolution of perator ${\mathsf
P}_{as}(|\phi_{mk}\rangle \langle\phi_{ml}| \otimes {\mathsf T}){\mathsf
P}_{as}$ depends on the Hamiltonian. The simplest imaginable end result is
${\mathsf P}_{as}(|\varphi_{mk}\rangle \langle\varphi_{ml}| \otimes {\mathsf
T}^{(m) \prime\prime}){\mathsf P}_{as}$, where $\varphi_{mk}$ is a state of a
system identical to ${\mathcal S}$ with separation status $D_m$, $D_m \cap
D^{(n)} = \emptyset$ for all $m$ and $n$, and ${\mathsf T}^{(m) \prime\prime}$
is a state of ${\mathcal A}^{(m)}$ with separation status $D^{(m)}$. Thus, end
state ${\mathsf T}_{\text{release}}$ that can be reconstructed from the formal
evolution is
\begin{equation}\label{release} {\mathsf T}_{\text{release}} =
\left(\sum_{m=1}^N\right) p_m \sum_{kl} \frac{c_{mk}c^*_{ml}}{p_m}
|\varphi_{mk}\rangle \langle\varphi_{ml}| \otimes {\mathsf T}^{(m)
\prime\prime} \otimes \prod_{r=1}^{N\setminus m}\otimes {\mathsf T}^{(r)}\ .
\end{equation} System ${\mathcal S}$ has a non-trivial separation status again
so that the release in each channel can be understood as an instance of
preparation for the composite ${\mathcal S} + {\mathcal A}^{(m)}$ and the
whole evolution as a random mixture of these single preparations. The formula
(\ref{release}) preserves the reduction.

The new rules that have been proposed as yet always correct the unitary formal
evolution determined by standard quantum mechanics by a reduction of the state
operator. The reduced state occurs in the formulas as the so-called "end
state". We assume that the time instant at which each end state formula
is valid is the time at which the detector gives its macroscopic signal. No
details of the time evolution to this end state is given. The end state itself
as well as any time evolution to it cannot be derived from quantum mechanics
but must simply be guessed and subjected to experimental checks. For the
question of detailed time evolution in particular, one had first to find some
observable aspects for it to show that the question does make sense.

An interesting case, which has some relevance to the end-time question and
which is a hybrid of the registration by non-absorbing and absorbing detector
is the Einstein, Podolsky and Rosen (EPR) experiment \cite{EPR}. We consider
Bohm's form of it \cite{bohm}. A spin-zero particle decays into two spin-1/2
ones, ${\mathcal S}_1$ and ${\mathcal S}_2$, that run in two opposite
directions. The state of composite ${\mathcal S}_1 \otimes {\mathcal S}_2$ is
then
\begin{equation}\label{EPRin} \frac{1}{\sqrt{2}}(|1_+\rangle\otimes
|2_-\rangle - |1_-\rangle\otimes |2_+\rangle)\ ,
\end{equation} where $|1_+\rangle$ is the spin-up state of ${\mathcal S}_1$,
etc. Finally, the spin of ${\mathcal S}_1$ is registered after some time at
which the particles ${\mathcal S}_1$ and ${\mathcal S}_2$ may be far away from
each other. Let the detector be a special case of fixed-signal one, as
described in Sec.\ 4.1.2. Hence, there are two sub-detectors, ${\mathcal
A}_1^{(+)}$ and ${\mathcal A}_1^{(-)}$ so that spin up of ${\mathcal S}_1$ is
associated with a signal from ${\mathcal A}_1^{(+)}$ and spin down with that
from ${\mathcal A}_1^{(-)}$. Let the state of ${\mathcal S}_1 + {\mathcal
A}_1^{(+)}$ containing the signal be ${\mathsf T}_1^{(+)\prime}$ and that of
${\mathcal S}_1 + {\mathcal A}_1^{(-)}$ be ${\mathsf
T}_1^{(-)\prime}$. Although ${\mathcal S}_1$ will be swallowed by the detector
(see Sec.\ 4.2.1), the left particle may remain accessible to
registration. Thus, our new rule is analogous to Eq.\ (\ref{release}):
\begin{equation}\label{EPR} {\mathsf T}_{\text{EPR}} = \frac{1}{2}
|2_+\rangle\langle 2_+| \otimes {\mathsf T}_1^{(+)} \otimes {\mathsf
T}_1^{(-)\prime} + \frac{1}{2} |2_-\rangle\langle 2_-| \otimes {\mathsf
T}_1^{(+)\prime} \otimes {\mathsf T}_1^{(-)}\ ,
\end{equation} where ${\mathsf T}_1^{(+)}$ and ${\mathsf T}_1^{(-)}$ are the
non-excited states of the corresponding sub-detectors. The state reduction
takes place at the time of the detector signal and has a {\em non-local}
character. We don't see any paradox in it. The only problem comes with the
generalisation to a relativistic theory: what is the correct simultaneity
plane? This problem has been solved by Keyser and Stodolsky \cite{KS}, see
also the discussion in \cite{survey}.

Formulas (\ref{endflex}), (\ref{endfix}) and (\ref{release}) can readily be
generalised to registration on a non-vector state (we avoid the use of the
term "mixed state" because it has different meaning for different
authors) ${\mathsf S}$ of ${\mathcal S}$. First, we have to decompose
${\mathsf S}$ into eigenstates of ${\mathsf O}$,
\begin{equation}\label{nonvec} {\mathsf S} = \sum_{nkml} S_{nkml}
|\phi_{nk}\rangle\langle\phi_{ml}|\ ;
\end{equation} the probability to register eigenvalue $o_k$ on ${\mathcal S}$
is
$$
p_m = \sum_k S_{mkmk}\ .
$$
Finally, because of the linearity of ${\mathsf U}$, everything we must do is
to replace the expressions in formulas (\ref{endflex}), (\ref{endfix}) and
(\ref{release}) as follows:
\begin{equation}\label{repl} \frac{c_{mk}c^*_{ml}}{p_m} \mapsto
\frac{S_{mkml}}{p_m}\ .
\end{equation}

Next, consider the case that the registered particle can miss the detectors
and enter into environment. We can use formula (\ref{endfix}) again by
modelling the part of the environment that the particle must join if it misses
the detector by one of the sub-detectors, ${\mathcal A}^{(N)}$, say.

This also explains the fact that Schr\"{o}dinger cat is never observed in
the linear superposition of life and death states. Indeed, in the case of
Schr\"{o}dinger cat, there is a radioactive substance releasing
alpha-particles and a detector of alpha-particles, the signal of which leads
to the death of the cat. Then, we can decompose the state of an alpha-particle
into that of it being in the nucleus or of being released and missing the
detector and that of hitting the detector, so that the above analysis is
applicable.

\subsubsection{Registration of composite systems} Formulas (\ref{endflex}) and
(\ref{endfix}) were obtained for registrations of one-particle systems. This
section will generalise them to many-particle ones. Composite systems can be
classified into {\em bound} and {\em unbound}. Bound systems such as atoms and
molecules can be dealt with in an analogous way as particles. The only change
is that map ${\mathsf P}_{as}$ is more complicated. Then, formulas
(\ref{endflex}) and (\ref{endfix}) are valid for bounded composite
systems. Unbounded composite systems are different. A system ${\mathcal S}$
that contains $K$ particles can excite more detectors simultaneously, at most
$K$ detectors.

Generalisation to such systems is not completely straightforward because it
must achieve, on the one hand, that there can be some non-trivial correlations
between the signals from different detectors and, on the other, that the
detectors are never in a linear superposition of their different signals,
which in turn erases some correlations between different detectors. Of course,
for one-particle systems, signals of different detectors are always
anti-correlated in a trivial way. Non-trivial correlations that can emerge for
unbounded many-particle systems are e.g.\ Hanbury-Brown-Twiss (HBT) ones
\cite{HBT} or Eistein-Podolski-Rosen (EPR) ones. Let us start with HBT effect.

In the original experiment by Hanbury Brown and Twiss, two photomultiplier
tubes separated by about 6 m distance, were aimed at the star Sirius. An
interference effect was observed between the two intensities, revealing a
positive correlation between the two signals. Hanbury Brown and Twiss used the
interference signal to determine the angular size of Sirius. The theory of the
effect \cite{fano} studies a model in which the signal consists of two photons
that impinge simultaneously on two detectors. Our strategy will be to
construct a non-relativistic model of Hanbury Brown and Twiss effect following
closely Fano's ideas \cite{fano} and try then to modify it similarly as the
BCL model has been modified for the case of one-particle systems.

Let us limit ourselves to ${\mathcal S} = {\mathcal S}_1 + {\mathcal S}_2$
consisting of two bosons, $K = 2$, with Hilbert spaces ${\mathbf H}_1$ and
${\mathbf H}_2$. To simplify further, let the registered observable be
${\mathsf O}_1 + {\mathsf O}_2$, ${\mathsf O}_k$ having only two eigenvalues
$+1$ and $-1$ and eigenvectors $|k_+\rangle$ and $|k_-\rangle$, $k = 1,2$
satisfying
$$
{\mathsf O}_k |k_+\rangle = +|k_+\rangle\ ,\quad {\mathsf O}_k |k_-\rangle =
-|k_-\rangle\ .
$$
Let, moreover, the one-particle Hilbert spaces be two-dimensional,
i.e. vectors $|k_+\rangle$ and $|k_-\rangle$ form a basis of ${\mathbf
H}_k$. Let the projections onto these states be denoted by ${\mathsf P}_{k+}$
and ${\mathsf P}_{k-}$ so that we have:
\begin{equation}\label{proj1} {\mathsf P}_{k+}{\mathsf P}_{k+} = {\mathsf
P}_{k+}\ ,\quad {\mathsf P}_{k-}{\mathsf P}_{k-} = {\mathsf P}_{k-}\ ,\quad
{\mathsf P}_{k+}{\mathsf P}_{k-} = 0\ .
\end{equation} The generalisation to more particles of arbitrary kinds,
general observables and general Hilbert spaces is straightforward.

The Hilbert space ${\mathbf H}$ of the composite system has then basis
$\{|++\rangle,|--\rangle,|+-\rangle\}$, where
\begin{eqnarray*} |++\rangle &=& |1_+\rangle |2_+\rangle\ ,\\ |--\rangle &=&
|1_-\rangle |2_-\rangle\ ,\\ |+-\rangle &=& \frac{1}{\sqrt{2}}(|1_+\rangle
|2_-\rangle + |1_-\rangle |2_+\rangle)\ .
\end{eqnarray*} It is the basis formed by eigenvectors of ${\mathsf O}_1 +
{\mathsf O}_2$ with eigenvalues $2$, $-2$ and $0$, respectively. The
corresponding projections are
\begin{eqnarray*}\label{proj2} {\mathsf P}_{++} &=& {\mathsf P}_{1+}{\mathsf
P}_{2+}\ ,\\ {\mathsf P}_{--} &=& {\mathsf P}_{1-}{\mathsf P}_{2-}\ ,\\
{\mathsf P}_{+-} &=& {\mathsf P}_{1+}{\mathsf P}_{2-} + {\mathsf
P}_{1-}{\mathsf P}_{2+}\ .
\end{eqnarray*} It follows from Eq.\ (\ref{proj1}) that these are indeed
projections.

To calculate the correlation in a state ${\mathsf S}$ of system ${\mathcal S}$
between the values $\pm 1$ of any subsystem ${\mathcal S}_1$ or ${\mathcal
S}_2$, which is intended to model the correlation measured by Hanbury Brown
and Twiss, we need probability $p_+$ that eigenvalue $+1$ will be registered
at least on one subsystem and similarly $p_-$ for $-1$. These are given by
\begin{eqnarray*} p_+ &=& tr[{\mathsf S}({\mathsf P}_{++} + {\mathsf
P}_{+-})]\ ,\\ p_- &=& tr[{\mathsf S}({\mathsf P}_{--} + {\mathsf P}_{+-})]\ ,
\end{eqnarray*} respectively. If we define
$$
{\mathsf P}_+ = {\mathsf P}_{++} + {\mathsf P}_{+-}\ ,\quad {\mathsf P}_- =
{\mathsf P}_{--} + {\mathsf P}_{+-}\ ,
$$
we have
$$
{\mathsf P}_{+-} = {\mathsf P}_+ {\mathsf P}_-\ .
$$
The normalised correlation (see, e.g., \cite{BLM}, p.\ 50) is then given by
\begin{equation}\label{corr} C({\mathsf S}) = \frac{tr[{\mathsf S}{\mathsf
P}_+ {\mathsf P}_-] - tr[{\mathsf S}{\mathsf P}_+]tr[{\mathsf S}{\mathsf
P}_-]}{\sqrt{tr[{\mathsf S}{\mathsf P}_+] - (tr[{\mathsf S}{\mathsf
P}_+])^2}\sqrt{tr[{\mathsf S}{\mathsf P}_-] - (tr[{\mathsf S}{\mathsf
P}_-])^2}}\ .
\end{equation}

For example, let $|\Phi\rangle$ be a general vector state in $\mathbf H$:
$$
|\Phi\rangle = a|++\rangle + b|--\rangle + c|+-\rangle\ ,
$$
where $a$, $b$ and $c$ are complex numbers satisfying
$$
|a|^2 + |b|^2 + |c|^2 = 1\ .
$$
Then,
$$
C(\Phi) = -\frac{|a|^2 |b|^2}{\sqrt{(|a|^2 - |a|^4)(|b|^2 - |b|^4)}}\ .
$$
The correlation lies, in general, between $0$ and $-1$. The value $-1$ occurs
for $c= 0$, means the strong anti-correlation and is the standard (trivial)
case for one-particle systems.

Next, we construct a suitable detector. System ${\mathcal S}$ can be prepared
in vector state $|\Phi\rangle$ with separation status $D$ where then fields
and screens split the beam $B$ of single particles corresponding to
$|\Phi\rangle$ into two beams, $B_+$ and $B_-$, each corresponding to an
eigenvalue $\pm 1$ of observable ${\mathsf O}_1$ or ${\mathsf O}_2$. Let
detector ${\mathcal A}$ consist of two sub-detectors, ${\mathcal A}^{(+)}$
placed in the way of the beam $B_+$ and ${\mathcal A}^{(-)}$ placed in the way
of $B_-$ so that the signal of ${\mathcal A}^{(+)}$ registers eigenvalue $+1$
and that of ${\mathcal A}^{(+)}$ eigenvalue $-1$ on the registered particle
similarly as in our model of fixed signal detector in Sec.\ 4.1.2. Let the
Hilbert spaces of the sub-detectors be ${\mathbf H}_+$ and ${\mathbf H}_-$.

Let the sub-detectors be prepared in initial states $|{\mathcal
A}^{(+)}0\rangle$ and $|{\mathcal A}^{(-)}0\rangle$ with separation statuses
$D^{(+)}$ and $D^{(-)}$, $D^{(+)} \cap D^{(-)} = \emptyset$, $D \cap D^{(\pm)}
= \emptyset$. After the interaction between ${\mathcal S}$ and ${\mathcal A}$,
the following states are relevant: $|{\mathcal A}^{(+)}1\rangle \in {\mathsf
P}_{as}({\mathbf H}_1 \otimes {\mathbf H}_+)$, $|{\mathcal A}^{(-)}1\rangle
\in {\mathsf P}_{as}({\mathbf H}_1 \otimes {\mathbf H}_-)$, $|{\mathcal
A}^{(+)}2\rangle \in {\mathsf P}_{as}({\mathbf H}_2 \otimes {\mathbf H}_+)$,
$|{\mathcal A}^{(-)}2\rangle \in {\mathsf P}_{as}({\mathbf H}_2 \otimes
{\mathbf H}_-)$, $|{\mathcal A}^{(+)}12\rangle \in {\mathsf P}_{as}({\mathbf
H}_1 \otimes {\mathbf H}_2 \otimes {\mathbf H}_+)$ and $|{\mathcal
A}^{(-)}12\rangle \in {\mathsf P}_{as}({\mathbf H}_1 \otimes {\mathbf H}_2
\otimes {\mathbf H}_-)$. These states describe one or two of the particles
being swallowed by one of the sub-detectors, they are associated with changes
of their separation status and include detector signals.

Finally, to register ${\mathsf O}_1 + {\mathsf O}_2$, the measurement coupling
${\mathsf U}$ must satisfy
\begin{eqnarray}\label{U++} {\mathsf U}{\mathsf P}_{as}(|++\rangle \otimes
|{\mathcal A}^{(+)}0\rangle \otimes |{\mathcal A}^{(-)}0\rangle) &=& {\mathsf
P}_{as}(|{\mathcal A}^{(+)}12\rangle \otimes |{\mathcal A}^{(-)}0\rangle)\ ,
\\ \label{U--} {\mathsf U}{\mathsf P}_{as}(|--\rangle \otimes |{\mathcal
A}^{(+)}0\rangle \otimes |{\mathcal A}^{(-)}0\rangle) &=& {\mathsf
P}_{as}(|{\mathcal A}^{(+)}0\rangle \otimes |{\mathcal A}^{(-)}12\rangle)\ ,
\\ \label{U+-} {\mathsf U}{\mathsf P}_{as}(|+-\rangle \otimes |{\mathcal
A}^{(+)}0\rangle \otimes |{\mathcal A}^{(-)}0\rangle) &=& {\mathsf
P}_{as}(|{\mathcal A}^{(+)}1\rangle \otimes |{\mathcal A}^{(-)}2\rangle)\ .
\end{eqnarray} Observe that operator ${\mathsf P}_{as}$ also exchanges
particles 1 and 2, which is a non-trivial operation on the right-hand side of
Eq.\ (\ref{U+-}).

Eqs.\ (\ref{U++}), (\ref{U--}) and (\ref{U+-}) describe the formal evolution
defining the three channels of the measurement. Each channel leads to the
composite signal due to a registration of one copy of system ${\mathcal
S}$. Thus, it can include signals of two detectors (Eq.\ (\ref{U+-})).

The formal evolution of state $\Phi$ would yield for the end state of the
system ${\mathcal S} + {\mathcal A}$:
\begin{multline}\label{end0a} {\mathsf U}{\mathsf J}(|\Phi\rangle \otimes
|{\mathcal A}^{(+)}0\rangle \otimes |{\mathcal A}^{(-)}0\rangle) = a{\mathsf
J}(|{\mathcal A}^{(+)}12\rangle \otimes |{\mathcal A}^{(-)}0\rangle) \\ +
b{\mathsf J}(|{\mathcal A}^{(+)}0\rangle \otimes |{\mathcal A}^{(-)}12\rangle)
+ c{\mathsf J}(|{\mathcal A}^{(+)}1\rangle \otimes |{\mathcal
A}^{(-)}2\rangle)\ .
\end{multline}

According to our theory, this state must be reduced to a gemenge with
component states, each of them corresponding to a single channel. Thus, the
correct end state ${\mathsf T}_{\text{comp}}$ of the whole system ${\mathcal
S} + {\mathcal A}$ after the measurement process described above is
\begin{multline}\label{end1a} {\mathsf T}_{\text{comp}} = |a|^2{\mathsf
J}(|{\mathcal A}^{(+)}12\rangle\langle{\mathcal A}^{(+)}12|) \otimes
|{\mathcal A}^{(-)}0\rangle\langle{\mathcal A}^{(-)}0| \\ (+) |b|^2 |{\mathcal
A}^{(+)}0\rangle\langle{\mathcal A}^{(+)}0| \otimes {\mathsf J}(|{\mathcal
A}^{(-)}12\rangle\langle{\mathcal A}^{(-)}12|) \\ (+) |c|^2 {\mathsf
J}\left(|{\mathcal A}^{(+)}1\rangle\langle{\mathcal A}^{(+)}1| \otimes
|{\mathcal A}^{(-)}2\rangle\langle {\mathcal A}^{(-)}2|\right)\ .
\end{multline} We assume that formula (\ref{end1a}) describes a special case
of the registration of many-particle systems by many detectors and that it
illustrates a method that can be used for more general cases. State ${\mathsf
T}_{\text{comp}}$ is an operator on ${\mathbf H} \otimes {\mathbf H}_+ \otimes
{\mathbf H}_-$ and it is a convex combination of three states each on a
different subspace of it. These three states are obtained by reconstruction
from the corresponding results of formal evolution in accordance with the
separation statuses. For example, the formal evolution gives for the first
state
$$
{\mathsf J}(|{\mathcal A}^{(+)}12\rangle\langle{\mathcal A}^{(+)}12| \otimes
|{\mathcal A}^{(-)}0\rangle\langle{\mathcal A}^{(-)}0|)\ ,
$$
but both particles are inside ${\mathcal A}^{(+)}$ and are, together with
${\mathcal A}^{(+)}$, separated from ${\mathcal A}^{(-)}$.

One can see that the pair of sub-detectors is in a well-defined signal state
after each individual registration on ${\mathcal S}$ and, at the same time,
the correlation contained in state $|\Phi\rangle$ that models the HTB
correlation is preserved and can be read off the signals of the
sub-detectors. This is of course due to the fact that HTB correlation is a
function of the absolute values $|a|$, $|b|$ and $|c|$, none of which is
erased by reduction of Eq.\ (\ref{end0a}) to Eq.\ (\ref{end1a}), while the
extra correlations due to the linear superposition depend on mixed products
such as $ab^*$ etc.

A different but analogous case is the EPR experiment. The composite system of
two fermions ${\mathcal S}_1$ and ${\mathcal S}_2$ is in initial state
(\ref{EPRin}). The detector consists of four sub-detectors, ${\mathcal
A}^{(+)}_1$, ${\mathcal A}^{(-)}_1$, ${\mathcal A}^{(+)}_2$ and ${\mathcal
A}^{(-)}_2$, where the first pair interacts only with ${\mathcal S}_1$ and the
second only with ${\mathcal S}_2$. The initial states of the sub-detectors are
${\mathsf T}^{(\pm)}_k$. The symbol ${\mathsf T}^{(\pm)\prime}_k$ denotes the
state of system ${\mathcal A}^{(\pm)}_k + {\mathcal S}_k$ in which the
sub-detector ${\mathcal A}^{(\pm)}_k$ swallows particle ${\mathcal S}_k$ and
sends its signal. Procedure analogous to that leading to formula (\ref{end1a})
will now give for the end state
\begin{equation}\label{end1b} \frac{1}{2}{\mathsf T}^{(+)}_1 \otimes {\mathsf
T}^{(-)\prime}_1 \otimes {\mathsf T}^{(-)\prime}_2 \otimes {\mathsf
T}^{(+)\prime}_2 (+) \frac{1}{2}{\mathsf T}^{(+)\prime}_1 \otimes {\mathsf
T}^{(-)\prime}_1 \otimes {\mathsf T}^{(-)}_2 \otimes {\mathsf
T}^{(+)\prime}_2\ .
\end{equation} Again, EPR anti-correlation of the sub-detector signals is
preserved even if the quadruple of the sub-detectors is always in a
well-defined signal state at the end.

\subsection{Non-ideal detectors} Non-ideal detectors may be the natural and
dominating case, from the experimental point of view. If a non-ideal detector
${\mathcal A}$ is hit by a system ${\mathcal S}$, there is only probability $0
< \eta < 1$, the intrinsic efficiency, that it will give a signal. From the
theoretical point of view, they are important examples because our simple
method of channels does not work for them.

We restrict ourselves to flexible-signal detectors with possible signals
enumerated by $m = 1,\ldots,N$ and suppose that, in general, $\eta_m$ depends
on $m$. The other cases can be dealt with in an analogous way. Let again the
separation status of ${\mathcal A}$ be $D_{\mathcal A}$. If ${\mathcal S}$ is
prepared in an eigenstate of ${\mathsf O}$ with eigenvalue $o_m$ which
formally evolves to ${\mathcal S}$ being inside $D_{\mathcal A}$ with
certainty, then the probability that ${\mathcal A}$ signals is $\eta_m$ and
not 1. Thus, the condition of probability reproducibility is not satisfied in
this case. Instead, we introduce the notion of {\em approximate probability
reproducibility}. Its meaning is that the detector does register eigenvalue
$o_m$ on ${\mathcal S}$ if it gives $m$-th signal, but we don't know anything,
if it remains silent.

To construct a model of this situation, we must first modify Eq.\
(\ref{unitar}) that expresses the idea of probability reproducibility into
what expresses the approximate probability reproducibility (within standard
quantum mechanics):
\begin{equation}\label{unitar'} {\mathsf U}(\phi_{mk} \otimes \psi) = C_m^1
\varphi_{mk} \otimes \psi^1_m + C_m^0\phi'_{mk} \otimes \psi^0_m\ ,
\end{equation} where $\phi'_{mk}$ is a suitable time evolution of $\phi_{mk}$
into $D_{\mathcal A}$ and $\varphi_{mk}$ are states of ${\mathcal S}$, $\psi$
is the initial, $\psi^1_m$ the signal and $\psi^0_m$ a no-signal states of
${\mathcal A}$. These states satisfy orthogonality relations
$$
\langle \psi |\psi^1_m \rangle = 0\ ,\quad \langle \psi^1_m |\psi^1_n \rangle
= \delta_{mn}\ ,\quad \langle \psi^0_m |\psi^1_n \rangle = 0\ ,\quad \langle
\varphi_{mk} |\varphi_{ml}\rangle = \langle \phi'_{mk} |\phi'_{ml}\rangle =
\delta_{kl}\ .
$$
The coefficients $C_m^1$ and $C^0_m$ are related by
$$
|C_m^1|^2 + |C_m^0|^2 = 1\ ,\quad |C_m^1|^2 = \eta_m\ .
$$

Measurement coupling ${\mathsf U}$ commutes with ${\mathsf P}_{as}$ because
the Hamiltonian leaves ${\mathbf H}_{as}$ invariant and with ${\mathsf N}$
because it is a unitary map. We can, therefore, replace Eq.\ (\ref{unitar'})
by the corresponding formal evolution:
\begin{equation}\label{formev1} {\mathsf U}{\mathsf J}(\phi_{mk} \otimes \psi)
= C_m^1 {\mathsf J}(\varphi_{mk} \otimes \psi^1_m) + C_m^0 {\mathsf
J}(\phi'_{mk} \otimes \psi^0_m)\ .
\end{equation} This is not a channel because it is not the formal evolution of
an initial state into an end state with a single detector signal. Indeed, no
signal is also a macroscopically discernible detector state. We have to return
to the formal evolution that starts with general state $\phi$ of ${\mathcal
S}$:
\begin{multline}\label{formev2} {\mathsf U}{\mathsf
J}(|\phi\rangle\langle\phi| \otimes |\psi\rangle\langle\psi|){\mathsf
U}^\dagger = \sum_{mn}\sum_{kl}c_{mk}c^*_{nl}\Big(C_m^1 C^{1*}_n |{\mathsf
J}(\varphi_{mk} \otimes \psi^1_m)\rangle \langle{\mathsf J}(\varphi_{nl}
\otimes \psi^1_n)| \\ + C_m^1 C^{0*}_n |{\mathsf J}(\varphi_{mk} \otimes
\psi^1_m)\rangle \langle{\mathsf J}(\phi'_{nl} \otimes \psi^0_n)| + C_m^0
C^{1*}_n |{\mathsf J}(\phi'_{mk} \otimes \psi^0_m)\rangle \langle{\mathsf
J}(\varphi_{nl} \otimes \psi'_n)| \\ + C_m^0 C^{0*}_n |{\mathsf J}(\phi'_{mk}
\otimes \psi^0_m)\rangle \langle{\mathsf J}(\phi'_{nl} \otimes
\psi^0_n)|\Big)\ ,
\end{multline}

To obtain a correct end state of a non-ideal detector, we have to discard the
cross-terms between $\psi^1_m$ and $\psi^1_n$ and between $\psi^1_m$ and
$\psi^0_n$. This is a general method that works also in the case that there
are channels. The result is
\begin{multline}\label{nonidvec} {\mathsf T}_{\text{nonid1}} =
\left(\sum_{m=1}^N\right) p_m \eta_m \sum_{kl} \frac{c_{mk} c^*_{ml}}{p_m}
|{\mathsf J}(\varphi_{mk} \otimes \psi^1_m)\rangle \langle{\mathsf
J}(\varphi_{ml} \otimes \psi^1_m)| \\ (+) \sum_{mn}\sum_{kl} c_{mk}
c^*_{nl}C_m^0 C^{0*}_n |{\mathsf J}(\phi'_{mk} \otimes \psi^0_m)\rangle
\langle{\mathsf J}(\phi'_{nl} \otimes \psi^0_n)|\ .
\end{multline} This is not yet a practical formula because the detector is
always in a state with high entropy, which is not a vector state. Hence, the
initial state is $|\phi\rangle\langle\phi| \otimes {\mathsf T}$, and the end
state is
\begin{equation}\label{nonid} {\mathsf T}_{\text{nonid2}} =
\left(\sum_{m=1}^N\right) p_m \eta_m \sum_{kl} \frac{c_{mk} c^*_{ml}}{p_m}
{\mathsf T}^1_{mkl} (+) \sum_{mn}\sum_{kl} c_{mk} c^*_{nl}{\mathsf
T}^0_{mnkl}\ ,
\end{equation} where we have made the replacements
$$
|{\mathsf J}(\varphi_{mk} \otimes \psi^1_m)\rangle \langle{\mathsf
J}(\varphi_{ml} \otimes \psi^1_m)| \mapsto {\mathsf T}^1_{mkl}
$$
and
$$
C_m^0 C^{0*}_n|{\mathsf J}(\phi'_{mk} \otimes \psi^0_m)\rangle \langle{\mathsf
J}(\phi'_{nl} \otimes \psi^0_n)| \mapsto {\mathsf T}^0_{mnkl}\ .
$$
Operators ${\mathsf T}^1_{mkl}$ and ${\mathsf T}^0_{mnkl}$ are determined by
the initial state and the formal evolution and satisfy the conditions:
\begin{description}
\item[A']
$$
tr[{\mathsf T}^1_{mkl}] = \delta_{kl}\ ,\quad tr[{\mathsf T}^0_{mnkl}] =
(1-\eta_m)\delta_{mn}\delta_{kl}\ .
$$
\item[B'] For any unit complex vector with components $c_k$,
$$
\sum_{kl} c_kc^*_l {\mathsf T}^1_{mkl}
$$
is a state operator on ${\mathbf H}_{as}$ and the state includes direct $m$-th
signal from the detector.
\item[C'] For any unit complex vector with components $c_{mk}$ (for all $m$
and $k$)
$$
\left(\sum_m p_m(1-\eta_m)\right)^{-1}\sum_{mn}\sum_{kl} c_{mk}
c^*_{nl}{\mathsf T}^0_{mnkl}
$$
is a state operator on ${\mathbf H}_{as}$ and the state includes no detector
signal from the detector.
\end{description}

\subsection{Particle tracks in detectors} Particle tracks in a Wilson chamber
look suspiciously similar to classical trajectories and have been an
interesting problem for quantum mechanics since the end of 1920's. There is
the classical paper by Mott \cite{mott} (see also \cite{heisenb}), which shows
by applying Schr\"{o}dinger equation that there is an overwhelming
probability of getting a second scattering event very close to the ray
pointing away from the decay centre through the location of a first scattering
event. A more rigorous calculation is given in \cite{DFT}, which uses the same
idea for a one-dimensional model. The initial situation is spherically
symmetric and the interaction between the alpha-particle and the detector also
is. Thus, the resulting state must also be spherically symmetric and not just
one radial track. A consequence of the linearity of Schr\"{o}dinger
equation then is that the end state is a linear superposition of all possible
radial tracks. A way to save one single radial track is the state reduction at
least for the first ionisation, which is apparently assumed tacitly. This
separation of state reduction and unitary evolution does not exactly
correspond to what is going on because we have in fact a chain of state
reductions with a unitary evolution in between.

In this section, we apply our theory to the problem, but we simplify it by
assuming, instead of the spherical symmetry, that the particle momentum has a
large average value $\langle \vec{p} \rangle$ and the detector has the plane
symmetry with the plane being perpendicular to $\langle \vec{p} \rangle$.

The registration model studied in subsection 4.1.2 can be characterised as a
single transversal layer of detectors: each beam is registered once. What we
now have can be viewed as an arrangement of many transversal detector layers:
one beam passes through all layers successively causing a multiple
registration. Examples of such arrangements are cloud chambers or MWPC
telescopes for particle tracking \cite{leo}. The latter is a stack of the
so-called multiwire proportional chambers (MWPC) so that the resulting system
of electronic signals contains the information about a particle track. Here,
we restrict ourselves to cloud chambers, but the generalisation needed to
describe MWPC telescopes does not seem difficult.

Then, a model of a Wilson chamber is a system of sub-detectors ${\mathcal
A}^{(nk)}$, where $n$ distinguishes different transversal layers and $k$
different sub-detectors in each such layer. Let the space occupied by
${\mathcal A}^{(nk)}$ be $D^{(nk)}$ and let it be at the same time its
separation status. We shall assume that $D^{(nk)}$ are small cubes with edge
$d$ that is approximately equal to the diameter of the resulting clouds in the
Wilson chamber. We denote the $n$-th layer by ${\mathcal A}^{(n)}$ so that
${\mathcal A}^{(n)} = \cup_{k=1}^N{\mathcal A}^{(nk)}$. To simplify the
subsequent analysis, we assume that coordinates can be chosen in a
neighbourhood of ${\mathcal A}^{(n)}$ so that each $D^{(nk)}$ in the
neighbourhood can be described by
$$
x^1 \in (u_k^1,u_k^1+d)\ ,\quad x^2 \in (u_k^2,u_k^2 +d)\ ,\quad x^3 \in
(u_n^3,u_n^3+d)\ .
$$

The observable ${\mathsf O}^{(n)}$ that is registered by each layer ${\mathcal
A}^{(n)}$ is equivalent to the position within the cubes. The eigenfunctions
and eigenvalues are
$$
{\mathsf O}^{(n)}\phi^{(nk)}_{l_1l_2l_3}(\vec{x}) = k
\phi^{(nk)}_{l_1l_2l_3}(\vec{x})\ ,
$$
where $\{l_1,l_2,l_3\}$ is a triple of integers that replaces the degeneration
index $l$,
$$
\phi^{(nk)}_{l_1l_2l_3}(\vec{x}) = d^{-3/2} \exp\left(\frac{2\pi
l_1i}{d}(x^1-u^1_k) + \frac{2\pi l_2i}{d}(x^2-u^2_k) + \frac{2\pi
l_3i}{d}(x^3-u^3_n)\right)
$$
for $\vec{x} \in D^{(nk)}$ and $\phi^{(nk)}_{l_1l_2l_3}(\vec{x}) = 0$
elsewhere.

The state ${\mathsf S}_n$ of ${\mathcal S}$ impinging on ${\mathcal A}^{(n)}$
can be defined as the state ${\mathcal S}$ would have after being released by
the layer ${\mathcal A}^{(n-1)}$. The interaction of ${\mathcal S}$ with
${\mathcal A}^{(n)}$ can then be described by formula (\ref{release}) with
replacement (\ref{repl}). The decomposition (\ref{nonvec}) must, of course,
use functions $\phi^{(nk)}_{l_1l_2l_3}$ instead of $\phi^{(n)}_k$ and the
support of $\varphi^{(nk)}_{l_1l_2l_3}$ is $D^{(nk)}$. The procedure can be
repeated for all $n$.

The first layer "chooses" one particular
$\varphi^{(1k)}_{l_1l_2l_3}$ with the support $D^{(1k)}$ in each individual
act of registration even in the case that the state arriving at it is a plane
wave. Hence, the "choice" in the next layer is already strongly
limited. In this way, a straight particle track of width $d$ results during
each individual multiple registration. Formally, of course, the resulting
state of ${\mathcal S}$ is a gemenge of all such straight tracks, which would
have the plane symmetry if the original wave arriving at the detector stack
were a plane wave.

\section{Changes of separation status \\ in scattering processes} It is the
existence of separation-status change that allows us to choose the gemenge
form, such as Eq.\ (\ref{endflex}), of the end states so that the theory
agrees with the observational fact of objectification. However,
separation-status changes can also occur in processes that have nothing to do
with registrations. Must there be any reduction to gemenge form in such
processes, too?

To study this question, let us restrict ourselves to a scattering of a
microsystem by a macroscopic target and observe that there can then be
separation status changes, one when the system enters the target and other
when it is released. First, let us consider no-entanglement processes such as
the scattering of electrons on a crystal of graphite with a resulting
interference pattern \cite{DG} or the splitting of a laser beam by a
down-conversion process in a crystal of KNbO$_3$ (see, e.g., Ref.\
\cite{MW}). No-entanglement processes can be described by the following
model. Let the initial state of the target ${\mathcal A}$ be $\mathsf T$ with
separation status $D_{\mathcal A}$ and that of the microsystem ${\mathcal S}$
be $\phi$ with separation status $D_1$, $D_1 \cap D_{\mathcal A} =
\emptyset$. Let there be two subsequent changes of separation status of
${\mathcal S}$: first, it is swallowed by ${\mathcal A}$ in $D_{\mathcal A}$
and, second, it is released by ${\mathcal A}$ in state $\varphi$ with
separation status $D_2$, $D_2 \cap D_{\mathcal A} = \emptyset$. We assume that
the end state of the target, ${\mathsf T}'$, is independent of $\phi$ and that
we have a unitary evolution:
$$
|\phi\rangle\langle\phi| \otimes {\mathsf T}
\mapsto|\varphi\rangle\langle\varphi| \otimes {\mathsf T}'\ ,
$$
which can be reconstructed from the formal evolution because the systems are
separated initially and finally. The two systems are not entangled by their
interaction, hence there is no necessity to divide the resulting correlations
between ${\mathcal S}$ and ${\mathcal A}$ in what survives and what is
erased. The end state is in fact of the form (\ref{release}): it has a trivial
gemenge structure.

Another example of this situation is a particle prepared in a cavity $D$ with
imperfect vacuum. We can model this situation in the above way and so in
effect suppose that the particle has separation status $D$.

A more interesting case is an {\em entanglement scattering} during which two
subsequent changes of separation status of the scattered particle also
occur. The scattering of neutrons on spin waves in ferromagnets or ionising an
atom of an ideal gas in a vessel are examples. Let microsystem ${\mathcal S}$
in initial state $\phi$ with separation status $D$ be scattered by a
macrosystem ${\mathcal A}$ in initial state ${\mathsf T}$ with separation
status $D_{\mathcal A}$, $D \cap D_{\mathcal A} = \emptyset$. For simplicity,
we assume that the formal evolution leads to supp$\phi \subset D_{\mathcal A}$
at some time $t_{\text{scatt}}$. Then, $t_{\text{scatt}}$ is not uniquely
determined but the subsequent calculations are valid for any possible choice
of it. A more general situation can be dealt with by the method applied in the
case of a microsystem that can miss a detector.

The experimental arrangement determines two Hilbert spaces ${\mathbf H}$ and
${\mathbf H}_{\mathcal A}$ and unitary map
\begin{equation}\label{scattU} {\mathsf U} : {\mathbf H} \otimes {\mathbf
H}_{\mathcal A} \mapsto {\mathbf H} \otimes {\mathbf H}_{\mathcal A}
\end{equation} describing the interaction according to standard quantum
mechanics.

The experimental arrangement studied in the previous section also determined a
basis $\{\phi_{mk}\}$ of ${\mathbf H}$, namely the eigenvectors of registered
observable ${\mathsf O}$ as well as sets of states $\{\varphi_{m k}\}$ in
${\mathbf H}$ and $\{\psi_m\}$ of ${\mathbf H}_{\mathcal A}$. This together
with the assumption that ${\mathcal A}$ measures ${\mathsf O}$ (with exact or
approximate probability reproducibility) restricted the possible ${\mathsf
U}$. These particular properties enabled us to choose a unique gemenge
structure for the end state. The question is how any gemenge form of the end
result can be even formally well-defined for processes described by Eq.\
(\ref{scattU}), where the physical situation does not determine any such
special sets of states.

To be able to give an account of the situation, let us first introduce the
formal evolution ${\mathsf U}_f$ on ${\mathsf P}_{as}({\mathbf H} \otimes
{\mathbf H}_{\mathcal A})$, from which ${\mathsf U}$ can be
reconstructed. Second, we decompose map ${\mathsf U}_f$ into two steps,
${\mathsf U}_f = {\mathsf U}_{f2} \circ {\mathsf U}_{f1}$, where ${\mathsf
U}_{f1}$ develops up to $t_{\text{scatt}}$ and ${\mathsf U}_{f2}$ further from
$t_{\text{scatt}}$.

Then, the correct intermediate state ${\mathsf T}_{\text{interm}}$ at
$t_{\text{scatt}}$ is
$$
{\mathsf T}_{\text{interm}} = {\mathsf N}({\mathsf U}_{f1}{\mathsf
P}_{as}(|\phi\rangle\langle \phi| \otimes {\mathsf T}){\mathsf P}_{as}{\mathsf
U}_{f1})\ .
$$
Indeed, there is no macroscopic signal from ${\mathcal A}$, only some
microscopic degrees of freedom of ${\mathcal A}$ change due to the interaction
${\mathsf U}_{f1}$. The overwhelming part of the degrees of freedom of
${\mathcal A}$ remains intact and just serve as a background of the
process. Thus, even if there is a separation status change, there is no
necessity for reduction: one can say that there is only one channel.

Further evolution is given by ${\mathsf U}_{f2}$ supplemented by
reconstruction of the states in ${\mathbf H}$ and ${\mathbf H}_{\mathcal A}$
as ${\mathcal S}$ is released by ${\mathcal A}$, and we simply obtain: the
formula
\begin{equation}\label{scatt} {\mathsf T}_{\text{end}} = {\mathsf
U}(|\phi\rangle\langle\phi| \otimes {\mathsf T}_0){\mathsf U}^\dagger
\end{equation} of standard quantum mechanics remains valid. Formula
(\ref{scatt}) makes clear that a separation status change need not cause any
reduction.

\section{Conclusion} Standard quantum mechanics does not contain rules
governing changes of separation status. We have utilised this opportunity to
construct the missing rule so that it satisfies the objectification
requirement.

Sec. 3.2 has introduced a new technical tool, the formal evolution, that
enables us to study changes of separation status in detail. In rigorous terms,
it describes the modification of kinematics due to separation status change on
the one hand and the role of Schr\"odinger equation in the process of
separation status change on the other.
 
Secs. 4 and 5 have discussed all possible kinds of experiments in which a
change of separation status occurs. A case by case analysis trying to take
into account the idiosyncrasy of each experiment and to isolate the relevant
features of its results has lead to formulas (\ref{endflex}), (\ref{endfix}),
(\ref{end1a}), (\ref{end1b}), (\ref{nonid}) and (\ref{scatt}). The real
purpose of the analysis however was to find a general rule so that each of the
formulas would be a special case of it. And indeed, now it is almost obvious
how the rule must read:
\par \vspace{.5cm} \noindent {\bf The Rule of Separation Status Change} {\it
Let microscopic system ${\mathcal S}$ be prepared in state ${\mathsf
T}_{\mathcal S}$ with separation status $D_{\mathcal S}$ and macroscopic
systems ${\mathcal A}$ in state ${\mathsf T}_{\mathcal A}$ with separation
status $D_{\mathcal A}$, where $D_{\mathcal S} \cap D_{\mathcal A} =
\emptyset$. The initial state is then ${\mathsf T}_{\mathcal S} \otimes
{\mathsf T}_{\mathcal A}$ according to Rule 2. Let the formal evolution
(defined in Sec.\ 3.2) describing the interaction between ${\mathcal S}$ and
${\mathcal A}$ lead to separation status change of ${\mathcal S}$. If there
are any macroscopic direct signals (defined in Sec.\ 4.1) from ${\mathcal A}$,
then the state of the composite ${\mathcal S} + {\mathcal A}$ given by the
formal evolution must be corrected by state reduction to the gemenge structure
(defined in \cite{hajicek2})
\begin{equation}\label{genrule} {\mathsf T}_{\text{end}} = \left(\sum_m\right)
p_m {\mathsf T}'_m\ ,
\end{equation} where each state ${\mathsf T}'_m$ includes only one (possibly
composite) direct signal from the whole detector. States ${\mathsf T}'_m$ of
${\mathcal S} + {\mathcal A}$ are determined by the formal evolution. The
state ${\mathsf T}_{\text{end}}$ refers then to any time after the signals.}
\par \vspace{.5cm} \noindent Hence, the evolution during a separation status
change brings three changes: first, the change of kinematic description
${\mathsf T}_{\mathcal S} \otimes {\mathsf T}_{\mathcal A} \mapsto {\mathsf
J}({\mathsf T}_{\mathcal S} \otimes {\mathsf T}_{\mathcal A})$, second, the
standard unitary evolution of state ${\mathsf J}({\mathsf T}_{\mathcal S}
\otimes {\mathsf T}_{\mathcal A})$, and third, the state reduction of the
evolved state into (\ref{genrule}). Afterwards, the state evolves unitarily
with a possible change of kinematics if ${\mathcal S}$ and ${\mathcal A}$
become separated again. Its form (the gemenge structure) is then uniquely
determined by detector signals. It is interesting to observe that the signals
result in a process of relaxation, in which the sensitive matter of the
detector approach its thermal equilibrium. This seems to be in accord with our
theory of classical states in \cite{hajicek1}.

A tenet adopted for the search of the Rule has been that corrections to
standard quantum mechanics ought to be the smallest possible changes required
just by the experiments. The Rule is of course guessed and not derived and
could yet be falsified in confrontation with further observational evidence
concerning different changes of separation status. It could also be further
extended, e.g., to describe how the postulated end states evolved in more
detail. However, for such an evolution, there does not seem to exist as yet
any experimental evidence to lead us. Let us emphasise that the clean
decomposition of a separation status change into three steps, viz.\ change of
kinematics, unitary evolution and state reduction, is just a method enabling a
mathematically well defined application of the Rule, but it is definitely not
a description of the time dependence of the real process.

Finally, we observe that The Reformed Quantum Mechanics returns to von
Neumann's "two kinds of dynamics" (see also \cite{hajicek2}) but
that its notion of state reduction differs from von Neumann's in two
points. First, it is less ad hoc because it is justified by the argument of
separation status change, which is logically independent from the proper
quantum measurement problem, and second, it is more specific because it
happens only in a detector and its form is determined by objective processes
inside the detector sensitive matter.

\subsection*{Acknowledgements} The author is indebted to Uwe-Jens Wiese for
drawing his attention to the Hanbury Brown and Twiss effect, to an anonymous
referee for mentioning the work by K. K. Wan and Ji\v{r}\'{\i} Tolar for
useful discussions.

\end{document}